\documentstyle [12pt]{article}
\newcommand{\be}{\begin{equation}}
\newcommand{\ee}{\end{equation}}
\newcommand{\ba}{\begin{eqnarray}}
\newcommand{\ea}{\end{eqnarray}}
\newcommand{\nn}{\nonumber}
\def\(#1){(\ref{#1})}
\def\hat #1{\mathaccent94{#1}}

\def\s#1{{\cal {#1}}}
\begin{document}

\begin{titlepage}
\begin{flushright}
LA--UR 94--3817
\end{flushright}

\vspace{7mm}

\begin{center}
{\Large\bf  Decoherence, re-coherence,
and the black hole information paradox}

\vspace{12mm}

{\large J.R. Anglin, R. Laflamme, and W.H. Zurek}

\vspace{3mm}
Theoretical~Astrophysics, T-6, Mail~Stop~B288,
Los~Alamos~National~Laboratory, Los~Alamos, NM~87545

\vspace{6mm}
{\large J.P. Paz}

\vspace{3mm}
Departamento~de~Fisica, FCEN, UBA, Pabellon~1, Ciudad~Universitaria,
1428~Buenos~Aires, Argentina

\end{center}

\vspace{4mm}

\begin{quote}

\hspace*{5mm} {\bf Abstract.} We analyze a system consisting of an
oscillator coupled to a field.  With the field traced out as an
environment, the oscillator loses coherence on a very short {\it
decoherence timescale}; but, on a much longer {\it relaxation
timescale}, predictably evolves into a unique, pure (ground) state.
This example of {\it re-coherence} has interesting implications both
for the interpretation of quantum theory and for the loss of
information during black hole evaporation.  We examine these
implications by investigating the intermediate and final states of the
quantum field, treated as an open system coupled to an unobserved
oscillator.

\noindent PACS number(s): 03.65.Bz, 04.70.Dy

\end{quote}
\end{titlepage}
\clearpage

\section{Introduction}

\subsection{Overview of decoherence}

Recent years have witnessed a significant increase of interest in the
process of {\it decoherence}\cite{d1,d3,d4,d5} --- the loss of
quantum coherence suffered by a quantum system in contact with an {\it
environment}.  An {\it environment} consists of degrees of freedom
which are coupled with, but not regarded as an integral part of the
system.  External variables are an obvious example, but even internal
degrees of freedom may constitute an environment, if they cannot be
followed by the observer.  Many quantum systems are therefore subject
to decoherence, and the phenomenon is thus widespread and important.

For example, it has been demonstrated that collective observables of
macroscopic quantum systems will lose quantum coherence very quickly by
this means.  This loss of coherence will proceed at very different
rates, depending on the initial state of the system.  Indeed, in the
simple models of quantum apparatus proposed to describe the process of
measurement, one can select an interaction Hamiltonian which commutes
with the observable of the recording apparatus\cite{d1}.  The
varying susceptibility of initial states to decoherence then allows one
to model apparent collapse of the wave packet.  The results can be
taken to imply that different outcomes of a measurement are all
present, but --- in the language of Everett\cite{everett} --- belong
to different branches of the universal wave vector.  Their simultaneous
detection is impossible: decoherence leads to {\it environment-induced
superselection rules}\cite{d1}, which effectively exclude a majority
of states from the Hilbert space of the open system.  In the context of
the Many Worlds Interpretation (MWI) of quantum mechanics,
environment-induced superselection supplies a preferred basis, which
selects the ``branches' into which the universal state vector is
``splitting''.  Decoherence can thus be thought of as a ``missing
link'' between the quantum universe and classical reality, in that
it provides the criterion for selection of preferred observables (such
as position) while supplying an effective definition for classicality,
as well as the rationale for the apparent ``collapse of the wave
packet''.

The process of decoherence has also been studied in the somewhat less
idealized, but still exactly solvable, model of an oscillator system
coupled to a quantum field representing the environment.  The evolution
of this system is known as {\it quantum Brownian
motion}\cite{b7,b8,b9}, and it also exhibits environment-induced
superselection\cite{b10}.  One can describe superselection in the case
of quantum brownian motion by appealing to the {\it predictability
sieve}\cite{b11} --- a formal implementation of the idea that the
preferred quantum states will be stable ({\it i.e.}, will minimize
entropy production) in spite of the coupling to the environment.  For
example, the predictability sieve selects {\it coherent states} as the
preferred states of an underdamped harmonic oscillator\cite{b12}.
Moreover, because decoherence occurs on timescales which are typically
much smaller than a system's dynamical timescales (or even timescales
associated with monitoring by an observer), one can convincingly argue
that similar ``nice'' states will be singled out by environment-induced
superselection in more realistic (and more complicated) situations.

Much of the discussion of the implications of the decoherence process
is based on the tacit assumption that ``decoherence is forever''; a
concern is often voiced that any sign of re-coherence would be
trouble\cite{r13}.  In this paper we explore the problem of
re-coherence, by exhibiting a model --- quantum Brownian motion with a
zero-temperature environment --- in which decoherence happens quickly,
but then gets ``undone'' slowly.  Since decoherence and information are
intimately related in quantum theory, it will turn out that this system
can serve as an instructive toy model for the information problem in
black hole evaporation.

\subsection{Outline of the calculation}

A brief and simplified preview of our calculation is in order.  We will
start our oscillator in a ``Schr\"{o}dinger Cat'' state, a
superposition of two well-separated coherent states.  The initial state
of the oscillator-field system will therefore be\footnote{In fact, we
will consider states that are only approximately direct products of
field and oscillator states; but the simplified outline presented in
this Section still describes the essential physics involved.}
\be
|\Psi_i\rangle = \left( c_+ |\psi_+\rangle  +
c_-|\psi_-\rangle\right)|0\rangle_{field}\;.\nn
\ee
(More complicated pure initial conditions can also be considered.)
Interaction with the environment will (approximately, and for a few
decoherence timescales) lead to
\ba\label{intro1}
\lefteqn{|\Psi_i\rangle = \left( c_+ |\psi_+\rangle
  + c_-|\psi_-\rangle\right)|0\rangle_{field} \to }
                   \hspace{.75 truein}\nn\\
& & c_+ |\psi_+(t)\rangle |\Phi_+(t)\rangle_{field}
 + c_-|\psi_-(t)\rangle|\Phi_-(t)\rangle_{field} = |\Psi(t)\rangle\;,
\ea
\vfill
\noindent where
\be\label{intro2}
\langle\Phi_+(t)|\Phi_-(t)\rangle << 1\;.
\ee
Hence the density matrix of the oscillator will be given by
\be\label{intro3}
\hat\rho = |c_+|^2\, |\psi_+(t)\rangle\langle\psi_+(t)|\,
    +\, |c_-|^2\, |\psi_-(t)\rangle\langle\psi_-(t)|\,
      + \s{O}\bigl(\langle\Phi_+(t)|\Phi_-(t)\rangle\bigr)\;.
\ee
Thus, at this stage one might feel justified in ``declaring victory'':
the decoherence has happened, as the form of the density matrix of the
system demonstrates.

However, our system is a {\it damped} harmonic oscillator.  In contact
with the vacuum of a quantum field, it will slowly (on the relaxation
timescale) approach the unique ground state {\it regardless of the
initial state}.  Hence, after a sufficiently long time,
$|\Psi(t)\rangle$ will approach
\ba\label{intro4}
|\Psi(\infty)\rangle &\equiv c_+|0\rangle|\Phi_+\rangle
          + c_-|0\rangle|\Phi_-\rangle\nn\\
&= |0\rangle \bigl( c_+|\Phi_+\rangle + c_-|\Phi_-\rangle\bigr)\;,
\ea
where $|0\rangle$ is --- in our example --- the ground state of the
harmonic oscillator.  Thus, decoherence seems to have ``gone away'': it
did not prevent the state of the oscillator from re-cohering into a
unique, pure state.  Moreover, as a consequence of this {\it
recoherence} the environment (field) has been put into a very awkward,
{\it pure} ``Schr\"{o}dinger Cat'' state of its own!

Before this process has been completed, each of the two systems
involved (the oscillator or the field, with the other traced out as
unobserved) was in a mixed state, by virtue of the correlations between
them.  However, in the end these correlations have all disappeared.
Or, to put it more precisely, the oscillator-field correlations have
been ``used up'' to force the field into a highly non-trivial state.
This state has the property (as one can anticipate intuitively, and as
we shall prove in more detail below) that, in spite of its undeniable
purity, it appears to be mixed when explored by approximately local
measurements (limited in time and space to less than the duration of
the recoherence episode).  Thus, the information which appeared to be
``lost'' to observers who could access only one of the two systems
(namely, the field) eventually re-emerges, but in a very obscure and
hard-to-exhibit form.

\subsection{Analogy with black hole evaporation}

This sequence of events --- possible rapid loss of information, and its
eventual re-emergence after a long time, but in a form difficult to
decipher --- is analogous to another interesting and fundamental
process which has received a lot of attention in recent years: black
hole evaporation\cite{14,15,16,18}.  There, gravitational collapse
rapidly increases the entropy of the Universe by the (usually large)
difference between the entropy of the collapsing material and the final
entropy of the black hole\cite{19}, which is given (in bits) by the
area of its horizon measured in units of square Planck length.

The collapse seems to increase the entropy of the Universe, because the
inside of the black hole horizon is inaccessible to external
observers.  Furthermore, black hole evaporation puts the field outside
the horizon into a mixed state which, when analyzed layer by layer,
appears to contain approximately black body radiation, with entropy at
least as large as the entropy shed by the black hole.  There appears to
be no information in the emitted radiation.  So, when at the end of the
process the black hole is gone (as seems likely), the entropy of the
Universe is larger by at least the entropy increase which occurred
during the collapse.  This process can be analyzed in some detail in
the case of the Witten black hole in $1+1$ dimensional
spacetime\cite{16,18}.  Results in this case are usually cited (in
spite of the facts that the analogy with the $3+1$ dimensional case is
only partial, and that the calculation cannot really be carried out for
the time when the black hole remnant ultimately disappears) as evidence
for the {\it black hole information paradox}\cite{bhip}.  This
paradox is that the fundamental equations of gravitation and field
theory seem to imply an irreversible increase of entropy in the process
of collapse, despite the fact that they are themselves reversible.
Reversibility cannot, of course, be established in this case simply by
appealing to the dynamics, since Einstein's theory necessarily predicts
a singularity inside the horizon, where the known laws of physics,
including general relativity, are thought to fail.  It is easy to
imagine that the reversibility of the the whole process is an early
victim in this failure\cite{failure}.

The decoherence {\it cum} re-coherence process described and analyzed
in this paper supports an alternative view.  One can imagine that the
information lost beyond the black hole horizon eventually re-emerges,
but not in any obvious form such that it could be detected by looking
at ``natural'' observables (anything reasonably local, or at least
confined to finite shells of the radiation emitted by the black hole).
Rather, the information is re-emitted in a horribly ``scrambled up''
manner, where the state left behind after the black hole evaporates can
still be pure (or at least no more mixed than the pre-collapse state),
but this purity can only be revealed by measurement of some
uncompromisingly {\it global} observable, which coherently and
simultaneously samples {\it all} of the emitted quanta.

Motivated by this analogy, we shall exhibit such global observables,
which can be computed exactly in our system by virtue of its
linearity.  It should of course be emphasized that this linearity that
allows our model to be exactly solvable also makes it a rather distant
analogue for the black hole evaporation process.  It is precisely the
inherent nonlinearity of general relativity which is responsible for
the central singularity, the event horizon, the ``no hair'' theorems,
and therefore for the unique value of the black hole entropy.
Nevertheless, the complexity of the pure ``global states'' of the field
generated in our simple example suggests how the information can be
preserved but remain ``hidden'', and thus, suggests a possible
resolution of the black hole information paradox.

\section{The calculation}

\subsection{The model}

Our model consists of a simple harmonic oscillator $Q$ coupled to a
massless scalar field $\phi(x)$ in one dimension.  To ensure that the
energy is bounded below, we choose the following Lagrangian
\cite{b8}, with an ultraviolet cut-off:
\ba\label{Luz}
L &=& {M_0\over2} (\dot{Q}^2 - \Omega_0^2Q^2) +
{1\over2}\int_{-\infty}^\infty\!dx\,(\partial_t\phi)^2
                - (\partial_x\phi)^2\nn\\
& &\qquad\qquad - g Q \int_{-\infty}^\infty\!dx\,
                  F(x) \partial_t\phi(x)\;.
\ea
The coupling constant $g$ has dimensions such that we can define from
it the frequency $\gamma_0\equiv {g^2\over 4M_0}$.  Once renormalized,
this frequency will correspond to a relaxation timescale.  $F(x)$ is a
smearing function which implements the cut-off on the field-oscillator
interaction.  For our later convenience, we choose the particular form
\ba\label{FGam}
F &=& {1\over\pi}\int_0^\infty\!dk\,
           {\Gamma_0\over\sqrt{\Gamma_0^2 + k^2}}\cos kx\nn\\
&=& {\Gamma_0\over\pi}K_0(\Gamma_0 x)\;.\ea
$K_0(\Gamma_0 x)$ is the modified Bessel function of order zero, which
is concentrated within $x < \Gamma_0^{-1}$, and is a delta-function
representation in the limit $\Gamma_0\to\infty$.

We quantize our model by defining the Hamiltonian operator
\ba\label{Huz}
\hat H &=& {1\over2}\int_{-\infty}^\infty\!dx\,
       \left( [\hat\pi(x) + g  F(x) \hat Q]^2 +
              [\partial_x\hat\phi(x)]^2\right)\nn\\
& &\qquad + {1\over2}\left(M_0^{-1}\hat P^2
                     + M_0\Omega_0^2 \hat Q^2\right)\;.
\ea
Here we set $\hbar=c=1$, and introduce the convention that operators
have circumflex accents, while c-numbers do not.  $\hat P$ and
$\hat\pi(x)$ are the canonical momenta.

In fact, this model is unitarily equivalent to several other
much-studied systems, including even the free massless field.  In
writing \(Huz), therefore, we are really choosing how the Hilbert space
is to be divided into field and oscillator degrees of freedom.  Our
criteria for doing this in the way that leads to \(Huz) are that we
demand that an identifiable oscillator exists, that it be coupled
locally (save for some UV smearing) to the field at the origin, and
that the expected energies for direct product states of field and
oscillator are finite.  (The latter stipulation is needed in order for
weak coupling to imply small entanglement between oscillator and field
at low energies; it amounts to a demand for an ultraviolet cut-off.)

To see this unitary equivalence, and to proceed in our calculation, we
diagonalize \(Huz) by defining the following normal modes, for $\omega
>0$:
\ba\label{AB}
\hat\phi(x) &=& {1\over\sqrt\pi} \int_0^\infty\!d\omega\,
     \hat{A}_\omega u_\omega(x) + \hat{B}_\omega \sin\omega x\nn\\
\hat\pi(x) &=& {1\over\sqrt\pi}  \int_0^\infty\!d\omega\,
  \hat\Pi^A_\omega v_\omega(x) + \hat\Pi^B_\omega \sin\omega x\nn\\
\hat Q &=& {1\over\sqrt\pi}\int_0^\infty\!d\omega\,
                        \hat\Pi^A_\omega q(\omega)\nn\\
\hat P &=& {1\over\sqrt\pi}\int_0^\infty\!d\omega\,
                                   \hat{A}_\omega p(\omega)\;.
\ea

The mode functions $u_\omega$ and $v_\omega$, and the co-efficients
$q(\omega)$ and $p(\omega)$, are found by solving ordinary differential
equations derived by iterating the Heisenberg equations of motion.  One
obtains
\vfill
\ba\label{modes}
u_\omega(x) &=& C(\omega)\left(\left[1-{\omega^2\over\Omega_0^2} +
{2\gamma_0\Gamma_0\omega^2\over\Omega_0^2(\Gamma_0^2 +
\omega^2)}\right]\cos\omega x\right. \nn\\
& & \qquad+\left.\left.{4\gamma_0\Gamma_0^2\omega^2\over\pi\Omega_0^2
       \sqrt{\Gamma_0^2+\omega^2}}P\!\!\!\!\!\!
        \int_0^\infty\!{dk\over\sqrt{\Gamma_0^2 + k^2}}\,
         {\cos kx\over k^2-\omega^2}\right. \right)\nn\\
v_\omega(x) &=& u_\omega(x) + {4\gamma_0\Gamma_0 C(\omega)\over
                 \Omega_0^2 \sqrt{\Gamma_0^2 + \omega^2}} F(x)\nn\\
q(\omega) &=& -g C(\omega){\Gamma_0\over
                    M_0\Omega_0^2\sqrt{\Gamma_0^2 + \omega^2}}\nn\\
p(\omega) &=& - M_0\omega^2 q(\omega)\;.
\ea
We will soon see that $u_\omega(x)$ can be given in a much more
transparent form.

The important normalization co-efficient $C(\omega)$ is defined so
that
\ba\label{C}
C^2(\omega) &=& \left[\left(1-{\omega^2\over\Omega_0^2}
     +{2\gamma_0\Gamma_0\omega^2\over\Omega_0^2(\Gamma_0^2
      + \omega^2)}\right)^2
       +\left({2\gamma_0\Gamma_0^2\omega\over\omega^2(\Gamma_0^2
                + \omega^2)}\right)^2\right]^{-1}\nn\\
&\equiv&{\Omega_0^4(\Gamma_0^2+\omega^2)\over[\omega^2+\Gamma^2]
    [\omega^2-(\Omega+i\gamma)^2][\omega^2-(\Omega-i\gamma)^2]}\;.
\ea

The convenience of the cut-off scheme defined in \(FGam) lies in the
fact that $C^2(\omega)$ has only six simple poles, occurring in pairs
of equal magnitude.  This will make it easy to perform analytically
several integrals that appear in the time evolutions discussed in later
subsections.  In Eq.(6) the new quantities $\Gamma, \Omega, \gamma$ are
modified versions of the frequencies $\Gamma_0, \Omega_0, \gamma_0$,
renormalized by the interaction.  The renormalized frequencies may be
expressed in terms of the bare ones only through a cumbersome (although
analytically solvable) cubic equation.  It turns out, however, that if
we fix the renormalized parameters, which are the physically relevant
ones, then the corresponding bare parameters may be expressed
relatively simply:
\ba\label{barbare}
\Gamma_0 &=& \Gamma + 2\gamma\nn\\
\Omega_0^2 &=& {\Gamma\over\Gamma + 2\gamma}(\Omega^2 + \gamma^2)\nn\\
\gamma_0 &=& \gamma\left({\Gamma\over\Gamma + 2\gamma} +{\Omega^2 +
\gamma^2\over(\Gamma +2\gamma)^2}\right)\;.
\ea

{From} \(barbare) one can see that if $\Gamma >> \Omega >> \gamma$,
then the same inequalities will hold for the bare quantities as well,
and the proportional differences between the two sets of frequencies
will all be of order $({\gamma\over\Omega})^2$,
$({\Omega\over\Gamma})^2$, or ${\gamma\over\Gamma}$.  {\it The results
in this subsection 2.1 are all exact, and do not assume any particular
relationships between the three frequencies.}  In the remainder of this
paper, however, we will be interested in the case of extreme
underdamping, with high cut-off frequency.  We will therefore set
\ba\label{ratios}
\gamma &=& \epsilon \Omega\nn\\
\Omega &=& \epsilon \Gamma\;,
\ea
where $\epsilon$ is small.  We will also consider
${1\over\pi}\epsilon(1+\ln\epsilon)$ to be negligible.  Only under this
stronger assumption will the oscillator's final state be approximately
pure\cite{b8}.

The mode functions $u_\omega$ and $v_\omega$ are both even functions in
$x$.  (Since the oscillator couples only to the even modes of the
field, the odd mode functions are merely the usual sines.  In our
scenario, the odd modes will simply remain in their ground states
forever, and so we will not refer to them explicitly hereafter.)  For
$\omega$ far from $\Omega_0$, $u_\omega$ and $v_\omega$ are essentially
cosines; but they are distorted near the origin for $\omega$ close to
$\Omega_0$, as one might expect.  They possess several properties
analogous to the orthonormality of cosines:
\ba\label{props}
\int_{-\infty}^\infty\!dx\,v_\omega(x) u_{\omega'}(x) &=&
\pi\delta(\omega-\omega') + q(\omega) p(\omega')\nn\\
\int_0^\infty\!d\omega\,v_\omega(x) u_\omega(y)
      &=& {\pi\over2}[\delta(x-y) + \delta(x+y)]\nn\\
\int_0^\infty\!d\omega\, q(\omega)u_\omega(x)
        &=& \int_0^\infty\!d\omega\, p(\omega)v_\omega(x) = 0\nn\\
\int_0^\infty\!d\omega\, p(\omega) q(\omega) &=& -\pi\;.
\ea

\eject

These relations may all be verified by contour integration.  The
computation is made easier if we use \(barbare) to re-write \(modes) as
\ba\label{modes2}
u_\omega(x) &=& \int_{-\infty}^\infty\!dx'\,
          F(x-x')\tilde u_\omega(x')\nn\\
\tilde u_\omega(x) &=&
{C(\omega)\over\Omega_0^2\Gamma_0\sqrt{\Gamma_0^2+\omega^2}}
          \left([(\Omega_0^2-\omega^2)(\Gamma_0^2+\omega^2)
            + 2\gamma_0\Gamma_0\omega^2]\cos\omega x\right.\nn\\
& &\qquad\qquad\qquad\qquad\qquad\qquad \left.-
       2\gamma_0\Gamma_0^2\omega\sin\omega|x|\right)\\
&=& -{C(\omega)\over\Omega_0^2\Gamma_0\sqrt{\Gamma_0^2+\omega^2}}\nn\\
& &\;\times\,\Re\left([\omega-(\Omega+i\gamma)]
         [\omega+(\omega-i\gamma)][\omega-i\Gamma]
           [\omega+i(\Gamma + 2\gamma)] e^{i\omega|x|}\right)\;.\nn
\ea

Using \(props) we can invert the transformation \(AB), to obtain
\ba\label{ABinv}
\hat{A}_\omega &=& {1\over\sqrt\pi}\left[\int_{-\infty}^\infty\!dx\,
           \hat\phi(x) v_\omega(x) \; - q(\omega) \hat P\right]\nn\\
\hat\Pi^A_\omega &=& {1\over\sqrt\pi}\left[\int_{-\infty}^\infty\!dx\,
         \hat\pi(x) u_\omega(x) \; - p(\omega) \hat Q \right]\;.
\ea

It is then straightforward to verify from the standard commutation
relations of $\hat\phi$ and $\hat\pi$, and $\hat Q$ and $\hat P$, that
\(ABinv) implies the canonical relations
\ba\label{CCR}
[{\hat A}_\omega, {\hat A}_{\omega'}] &=& 0 \nn \\  ~
[{\hat\Pi^A}_\omega, {\hat\Pi^A}_{\omega'}] &=& 0 \nn \\ ~
[{\hat A}_\omega, {\hat\Pi^A}_{\omega'}] &=& i\delta(\omega-\omega')\;.
\ea

Furthermore, we have
\be\label{Hdiag}
\hat H = {1\over2}\int_0^\infty\!dw\,\left((\hat\Pi^A_\omega)^2 +
\omega^2\hat{A}_\omega^2 + \right) + \hat H_{odd}\;,
\ee
where $\hat H_{odd}$ contains the unimportant odd mode operators
$B_\omega$ and $\Pi^B_\omega$.  This demonstrates that \(Huz) is indeed
equivalent to a free massless field.  It is of course obvious that any
quadratic model is equivalent to some spectrum of decoupled
oscillators; but the fact that the single oscillator simply disappears
like a drop in the continuous bucket of field modes, and does not alter
the spectral density at all, is not so trivial.

\subsection{Initial state}

Previous investigations of decoherence have typically considered direct
product initial states, of the form
\be\label{typinit}
|\Psi_i\rangle = |0\rangle_{field} |\psi_a\rangle_{oscillator}\;,
\ee
where $|0\rangle_{field}$ is the vacuum state of the free field, and
$|\psi_a\rangle$ is a bimodal oscillator state, such as
\be
|\psi_a\rangle = c_+ \, e^{-ia\hat P} |0\rangle_{osc} + c_- \,
         e^{ia\hat P} |0\rangle_{osc}\;,\nn
\ee
$|0\rangle_{osc}$ being the free oscillator ground state.   For
calculational convenience, we will instead use the initial state
\be\label{init}
|\Psi_i\rangle =  c_+ \, e^{-ia\hat P} |0\rangle + c_- \, e^{ia\hat P}
|0\rangle\;,
\ee
where $|0\rangle$ is the true, interacting ground state of the
field-oscillator system.

Since our objective in this paper is to study cases in which the
initial oscillator entropy is effectively zero, we must show that the
field-oscillator entanglement in the initial state \(init) leads only
to negligible initial entropy when the field is traced out.  Since we
have diagonalized the full Hamiltonian, the wave functional of the
interacting ground state $|0\rangle$ in the $\Pi^A_\omega$ variables is
simply a product of Gaussians:
\be\label{ground}
\langle \Pi^A|0\rangle = Z^{-1}
e^{-{1\over2}\int_0^\infty\!{d\omega\over\omega}\,(\Pi^A_\omega)^2}\;.
\ee
$Z$ is a normalization constant; a state $\langle\Pi^A_\omega|0\rangle$
is an eigenstate of the $\hat \Pi^A_\omega$ operators defined in
\(ABinv)\footnote{Both entities are of course only well-defined if we
consider the continuous and infinite spectrum of oscillators to be the
limit of a sequence of systems of finite numbers of oscillators.}.

We can readily obtain from \(ground) the corresponding reduced density
matrix  for the oscillator,
\be\label{RDM0}
\rho_0(Q,Q') = \int\!\s{D}\pi\, \langle Q, \pi|0\rangle\langle
0|Q',\pi\rangle\;,
\ee
where we write $\s{D}\pi$ for the measure to indicate that we want the
limit where the integral is continuously infinite dimensional.  Since
according to \(ABinv) the operators $\hat\Pi^A_\omega$ are linear
combinations of the operators $\hat\pi(x)$ and $\hat Q$, the states
$|\pi\rangle|Q\rangle$ are just different labellings of the states
$|\Pi^A_\omega\rangle$.  We therefore already have in equation
\(ground) the integrand of \(RDM0), and we need only now express the
path-like integral over $\pi$ in terms of the $\Pi^A_\omega$
variables.

We can deduce a straightforward way to do this from Equation \(ABinv).
We first change variables in \(RDM0) by substituting
\be
\Pi^A_\omega[\pi(x), Q] \to \bar\Pi^A_\omega[\pi(x)] -
{p(\omega)\over\sqrt\pi}Q\;,
\ee
in \(ground), where
\be
\bar\Pi^A_\omega \equiv {1\over\sqrt\pi}\int_{-\infty}^\infty\!dx\,
           u_\omega(x) \pi(x)\;.
\ee
Equation \(props) then implies
\be\label{alphabar}
\int_0^\infty\!d\omega\, q(\omega)\bar\Pi_\omega = 0\;.
\ee
To integrate over $\pi(x)$, then, we will integrate over
$\bar\Pi_\omega$, with a delta-function inserted to enforce \(alphabar)
and thus remove the $Q$ sector from the integral:
\ba\label{RDM1}
\rho_0(Q,Q') &=& \int\!d\lambda\,\int\!\s{D}\bar\Pi\,
e^{i{\lambda\over\sqrt\pi}\int_0^\infty\!d\omega\,
            q(\omega)\bar\Pi^A_\omega}e^{-{1\over2}
             \int_0^\infty\!{d\omega\over\omega}\,
              \left[(\bar\Pi^A_\omega - {p(\omega)\over\sqrt\pi}Q)^2
     + [(\bar\Pi^A_\omega - {p(\omega)\over\sqrt\pi}Q')^2\right]}\nn\\
&=& N e^{-{1\over4\pi} \int_0^\infty\!{d\omega\over\omega}\,
         p^2(\omega)(Q-Q')^2} e^{-{\pi\over4}[\int_0^\infty\!d\omega\,
             \omega    q^2(\omega)]^{-1} (Q+Q')^2}\nn\\
&\equiv& N e^{-{1\over2} M\Omega (Q^2 + Q'^2)}
                     e^{-\Delta M\Omega (Q-Q')^2}\;.
\ea
Here we introduce the renormalized mass $M$ and ground state
entanglement parameter $\Delta$, which with our choice of frequency
ratios \(ratios) are given by
\ba\label{MD}
M &=&  M_0\times\left(1 + {2\over\pi}{\gamma\over\Omega}
+\s{O}(\epsilon^2)\right)\nn\\
\Delta &=& -{1\over\pi}\epsilon(1+\ln\epsilon) +\s{O}((\epsilon^2)\;.
\ea

Since we have assumed (in order to have a pure state at late times)
that $\Delta$ is negligibly small, we now observe that \(RDM1) is
effectively equal, for all $Q, Q'$, to the density matrix of the (pure)
ground state of an oscillator with mass $M$ and natural frequency
$\Omega$.  Since the translation operators $e^{\pm ia\hat P}$ in
\(init) only shift $Q$ and $Q'$ in \(RDM1) by $\pm a$, it is also true
that the reduced density matrix formed from \(init) differs negligibly
from the one derived from \(typinit).  We have therefore shown that our
oscillator has negligible entropy, when the field is traced out of our
chosen initial state.  In fact, our choice of \(init) instead of
\(typinit) will have no significant effect on our results (because we
assume such weak coupling).

For the remainder of this paper, we will set $M\Omega = 1$, and assume
that, in the choice of units this implies, $a^2$ is large (of order
$\epsilon^{-1}$).  This will mean that, even though the oscillator is
very weakly coupled to the field, and has a dissipation timescale much
longer than the dynamical timescale, the two Gaussians that are
superposed in the initial state are far enough apart from each other
that significant decoherence will occur on a very short timescale.

\subsection{Oscillator evolution}

By using the transition matrix in the momentum representation for a
harmonic oscillator of frequency $\omega$, we easily obtain the wave
functional of the final state into which the initial state \(init)
evolves at time $t$:
\ba\label{Psit}
\langle\Pi^A ,\Pi^B |\Psi_t\rangle &=& N^{1\over2}
         e^{-{1\over2}\int_0^\infty\!{d\omega\over\omega}\,
           \left[(\Pi^A_\omega)^2 + i a^2{p^2(\omega)\over\pi}
                   \sin\omega t e^{-i\omega t}\right]}\nn\\
& &\qquad\times\left(c_+ e^{i{a\over\sqrt\pi}
          \int_0^\infty\!{d\omega\over\omega}\,p(\omega)
             e^{-i\omega t}} + c_- e^{-i{a\over\sqrt\pi}
           \int_0^\infty\!{d\omega\over\omega}\,p(\omega)
               e^{-i\omega t}}\right)\;.
\ea
We then use a technique like that employed in deriving \(RDM1) above to
obtain the reduced density matrix.

There will be four components:
\be\label{four}
\rho(Q,Q';t) = |c_+|^2\rho_{++} + |c_-|^2\rho_{--}
               + c_+c_-^*\rho_{+-} + c_- c_+^*\rho_{-+}\;.
\ee
We find that the `diagonal' components are given by
\ba\label{diag}
\rho_{\pm\pm}(Q,Q';t) &=& N e^{-{1\over2}\left([Q\mp a r(t)]^2
           + [Q'\mp a r(t)]^2 \right)}\nn\\
& &\qquad\times e^{\mp i a s(t) (Q-Q')}\nn\\
& &\qquad \times e^{- \Delta [Q-Q']^2} \;;
\ea
while the cross-terms are
\ba\label{cross}
\rho_{\pm\mp}(Q,Q';t)  &=& N e^{-{1\over2}\left([Q\mp a y(t)]^2
                    + [Q'\pm a y(t)]^2 \right)}\nn\\
& &\qquad\times e^{\mp i a z(t) (Q+Q')}\nn\\
& &\qquad \times e^{- \Delta [Q-Q'\mp 2 y(t)]^2}\nn\\
& &\qquad\times e^{- a^2\left((1+4\Delta)[1-y^2(t)] - z^2(t)\right)}\;.
\ea
Since $\Delta$ is negligible, we can ignore the last line of \(diag)
and the second-last line of \(cross).  This leaves each of the four
terms in $\rho(Q,Q';t)$ as a separated function of the form
\be\label{sep}
\rho_{\pm\pm'} =  \psi_{\pm\pm'}(Q)\psi^*_{\pm\pm'}(Q')\;.
\ee

The functions $r(t)$ and $s(t)$ are simple enough:
\ba\label{yz}
r(t) &=& e^{-\gamma t}[\cos\omega t - \epsilon \sin\omega t]
         +\s{O}(\epsilon^2)\nn\\
s(t) &=& e^{-\gamma t}[(1-{2\over\pi}\epsilon)\sin\Omega t + 2 \epsilon
        \cos\Omega t] - 2\epsilon e^{-\Gamma t} + \s{O}(\epsilon^2)\;.
\ea
It is clear from \(diag) and \(yz) that the diagonal terms describe
Gaussian wave packets performing weakly damped oscillations.

The functions $y(t)$ and $z(t)$, on the other hand, include some
exponential-integral terms:
\ba\label{uv}
y(t) &=& e^{-\gamma t}[(1-{2\over\pi}\epsilon-4\Delta)\cos\Omega t
         - 2\epsilon\sin\Omega t] -{2\over\pi}\epsilon
              + \s{O}(\epsilon^2)\nn\\
& &\qquad - {4\over\pi}\epsilon\Re\left((1+i{\Omega t\over2})
            e^{i\Omega t} [{\rm Ei}(-i\Omega t) + i\pi]\right)\nn\\
& &\qquad  + {2\over\pi}\epsilon\left(e^{\Gamma t} {\rm Ei}
        (-\Gamma t) + e^{-\Gamma t}{\rm Ei}(\Gamma t)\right)\nn\\
z(t) &=& e^{-\gamma t}[\sin\Omega t + \epsilon\cos\Omega t]
                           + \s{O}(\epsilon^2) \nn\\
& &\qquad -{2\over\pi}\epsilon \Im\left((1+i\Omega t)
           e^{i\Omega t}[{\rm Ei}(-i\Omega t) + i\pi]\right) \;.
\ea
(The exponential-integral functions of imaginary argument appear
together with $i\pi$ because we actually need a different branch of the
${\rm Ei}$ function than the standard one.)

The cross-terms are very rapidly suppressed, because the exponent of
the last term in \(cross),
\be\label{D}
D(t)\equiv a^2(1+4\Delta)[1-y^2(t)] - a^2z^2(t)\;,
\ee
grows on the cut-off timescale (see Figure 1), and $a^2$ is large.  We
emphasize that this decoherence occurs even in the extremely
underdamped limit where $\Delta \to 0$, and even when the initial state
is not an exact direct product.

The decohering factor $D(t)$ grows rapidly because at times much less
than a dynamical time $\Omega^{-1}$, $y(t)$ is approximately given by
\be\label{Dearly}
y(t) \sim 1 + {2\over\pi}\epsilon\left( e^{\Gamma t}
         {\rm Ei}(-\Gamma t) + e^{-\Gamma t}{\rm Ei}(\Gamma t)
             - 2{\rm C}_{Euler} - 2\ln(\Gamma t)\right)
                  + \s{O}(\epsilon^2)\;.
\ee
This function drops from unity on the cut-off timescale $\Gamma^{-1}$.
We can therefore see that $y(t)$ differs from $r(t)$, and $z(t)$
differs from $s(t)$, on the same timescale as $D(t)$ suppresses the
crossterms.  Hence there is a very short, very early time interval
during which the four wave functions $\psi_{\pm\pm'}$ appearing in
$\rho(Q,Q';t)$ are all significant and distinct.  Orthogonality,
however, as opposed to mere distinctness, is what will be important for
determining the eigenvalues and hence the entropy of the reduced
density matrix.  By the time $y(t)$ has diverged from $r(t)$ enough
that $\psi_{\pm\mp}$ are effectively orthogonal to $\psi_{\pm\pm}$,
$D(t)$ is already large, and $\psi_{\pm\mp}$ may be ignored.  We will
therefore be able to neglect the distinction between $\psi_{\pm\pm}$
and $\psi_{\pm\mp}$, and approximate the density matrix by the simpler,
two-state form
\ba\label{2simp}
\rho(Q,Q';t) &\doteq& \  |c_+|^2\, \psi_+(Q)\psi_+^*(Q')\
                 +\  |c_-|^2\, \psi_-(Q)\psi_-^*(Q') \nn\\
& &\qquad + \ e^{-D(t)}\,c_+c_-^*\,\psi_+(Q)\psi_-^*(Q')\nn\\
& &\qquad + \ e^{-D(t)}\,c_-c_+^*\,\psi_-(Q)\psi_+^*(Q')\;,
\ea
where
\be
\psi_{\pm}(Q) =\psi_{\pm\pm}(Q) \equiv N^{1\over2}
                 e^{-{1\over2}[Q\mp r(t)]^2 \mp i a s(t) Q}\;.
\ee

\subsection{Entropy evolution}

We can explicitly diagonalize $\hat\rho(t)$ in the approximation that
\(2simp) is valid by assuming that the eigenvector wave functions are
of the form

\be\label{philam}
\phi_\lambda(Q) = \sum_{\pm} A_{\pm}(\lambda) \psi_{\pm}(Q)\;.
\ee
We then solve for the co-efficients $A_{\pm}$ by requiring them to be
elements of the eigenvectors of the matrix
\be\label{matrix}
\s{M} \equiv \left(\begin{array}{cc}
|c_+|^2 + c_+c_-^* e^{-D(t) + a^2(r^2 + s^2)} & |c_+|^2 e^{-D(t)}
           + c_+c_-^* e^{-a^2(r^2 + s^2)}\\
|c_-|^2 e^{-D(t)} + c_-c_+^* e^{-a^2(r^2 + s^2)} & |c_-|^2
           + c_-c_+^* e^{-D(t) + a^2(r^2 + s^2)}
\end{array}\right)\;.
\ee
The quadratic characteristic equation for $\s{M}$ yields the two
eigenvalues
\be\label{ev}
\lambda_\pm = {1\over2}\left[ 1 \pm \sqrt{1 - 4 |c_+|^2 |c_-|^2
              (1 - e^{-2a^2e^{-2\gamma t}} - e^{-2D(t)})}\right]
                  + \s{O}(\epsilon^2)\;,
\ee
when we take advantage of the facts that $|c_+|^2 + |c_-|^2 = 1$,
$e^{-a^2(r^2+s^2)}\simeq e^{-a^2 e^{-2\gamma t}}$, and $e^{-D(t)}
e^{-a^2 e^{-2\gamma t}}\simeq 0$.

We now specialize to the case $c_+ = c_- ={1\over\sqrt2}$, where we
have

\be
\lambda_{\pm} \simeq {1\over2}\left(1 \pm [e^{-a^2e^{-2\gamma t}}
                + e^{-D(t)}]\right)\;,
\ee
since $e^{-a^2e^{-2\gamma t}} e^{-D(t)}$ is extremely small for any
$t$.

The entropy $S(t)$ for this case is plotted in Figure 2.  It initially
rises on the decoherence timescale from its initial negligible value to
$\ln 2$, where it persists for many dynamical times, before declining
on the dissipative timescale.  The re-establishment of the purity of
the oscillator state is clearly due to the fact that after a time on
the order of $\gamma^{-1}\ln a$ the two shifted gaussians have lost so
much amplitude that they begin to overlap and become
indistinguishable.  It then becomes less and less true that the
oscillator is in a mixture of two orthogonal states.  In the limit of
complete relaxation, the ground state is reached, and this is of course
a pure state.

We can assess the accuracy of the approximation \(2simp) by finding the
eigenvalues of $\hat\rho(t)$ using \(sep) instead.  In this case we
would have a fourth rank matrix in the analogue of \(matrix), and we
would find four eigenvalues.  Only two of these would be
non-negligible, however, and they would turn out to differ
insignificantly from \(ev).  The eigenvectors we would find by this
more accurate technique would differ somewhat, at very early times,
from the two $\phi_\lambda$ implied by \(philam) and \(matrix).  At
these early times, the more exact eigenvectors would also include some
non-vanishing amplitudes for the states represented by
$\psi_{\pm\mp}(Q)$, with their dependencies on $y(t)$ and $z(t)$
instead of $r(t)$ and $s(t)$.

\subsection{Field evolution}

We now wish to consider the massless scalar field as the observed
system, and to trace out the harmonic oscillator as unobserved.  Since
the initial state of the total system is pure, the non-zero eigenvalues
of the reduced density matrix of the field are the same as those of the
reduced density matrix of the oscillator.  The entropy of the field is
therefore the same as the oscillator entropy discussed in the preceding
subsection.  The problem that still remains, and which did not arise
for the oscillator, is that of assessing where in spacetime the
information associated with this entropy may be said to reside.

There are of course many possible definitions of the term
``information'', but for the purposes of this paper we shall consider
that the information problem will be solved by identifying the field
state as a simple mixture of states that can be created from the vacuum
by external sources.  The external sources will be functions in
spacetime, and the required information will be considered to reside in
the regions where these sources have support.

We will present the reduced density matrix $\hat R(t)$ for the field in
the basis of field operator eigenstates, and so we will need to
transform the field-oscillator state
$\langle\Pi^A_\omega,\Pi^B_\omega|\Psi_t\rangle$ of Equation \(Psit)
from the $\Pi^A_\omega, \Pi^B_\omega$ representation into the
$A_\omega, B_\omega$ representation.  We then invoke one of the
eigenvalue relations corresponding to Equation \(ABinv)
\be
A_\omega = {1\over\sqrt\pi}\left[\int_{-\infty}^\infty\!dx\,
            \phi(x) v_\omega(x) \; - q(\omega) P\right]\;,
\ee
and trace over $P$ to obtain the reduced density matrix
\be
R[\phi,\phi';t] = \int\!dP\, \langle A_\omega[\phi, P]
       |\Psi_t\rangle\langle\Psi_t| A_\omega[\phi',P]\rangle\;.
\ee

The reduced density matrix for the field will again be the sum of four
contributions,
\be\label{R4}
R[\phi,\phi';t] = |c_+|^2 R_{++} + |c_-|^2 R_{--} + c_+c_-^* R_{+-}
              + c_-c_+^*R_{-+}\;,
\ee
where
\ba\label{Rcomp}
R_{\pm\pm} &=&  \Psi_{\pm}[\phi;t]\Psi_{\pm}^*[\phi';t]
  e^{\left[\int_{-\infty}^\infty\!dx\,(\phi+\phi') L(x)\right]^2}\nn\\
R_{\pm\mp} &=&  \Psi_{\pm}[\phi;t]\Psi_{\mp}^*[\phi';t]
  e^{\left[\int_{-\infty}^\infty\!dx\,(\phi+\phi') L(x)\right]^2}\nn\\
& &\qquad\times e^{-a^2(r^2 + s^2)} e^{\pm a [i r(t) - s(t)]
          \int_{-\infty}^\infty\!dx\,(\phi+\phi') L(x) }\;.
\ea
Here we have defined
\ba\label{Lpsi}
L(x) &\equiv& {1\over\pi}\int_0^\infty\!d\omega\,
             \omega \,q(\omega) v_\omega(x)\nn\\
\Psi_{\pm}[\phi(x);t] &\equiv& \tilde Z^{-{1\over2}}\;
       e^{-{1\over2}\int_0^\infty\!d\omega\,{\omega\over\pi}
       \left( {a p(\omega)\over\omega}\sin\omega t + a q(\omega) z(t)
      - \int_{-\infty}^\infty\!dx\,\phi(x) v_\omega(x) \right)^2}\nn\\
& & \qquad\qquad\times\ e^{\pm ia
        \int_{-\infty}^\infty\!dx\, \phi(x) K(x,t)} \;.
\ea
The function $K(x,t)$ will be defined below.

As in \(2simp), we can simplify \(Rcomp) by approximating it as
\ba\label{Simp}
R_{\pm\pm} &\simeq& \Psi_{\pm}[\phi;t] \Psi_{\pm}^*[\phi';t]\nn\\
R_{\pm\mp} &\simeq& \Psi_{\pm}[\phi;t]\Psi_{\mp}^*[\phi';t]
            \times  e^{-a^2 e^{-2\gamma t}} \;.
\ea
This approximation can evidently lead to large errors in evaluating
expectation values of $\hat\phi(x)$ for $x$ close to the origin, where
$L(x)$ is non-negligible; but as with the analogous approximation in
\(2simp), it retains the actual behavior of the entropy very well.
Equation \(Simp) and the inner product
\be\label{inprod}
\langle\Psi_-(t)|\Psi_+(t)\rangle = e^{-D(t)}
\ee
imply that the eigenvalues of the density matrix \(R4) are again given
by \(ev), as should be the case.

In \(Simp), there are only two wave functionals $\Psi_{\pm}$ that
characterize the state of the field.  The states they represent may be
created by an external source, linearly coupled to both $\hat\phi$ and
$\hat\pi$:
\be\label{source}
|\Psi_{\pm}(t)\rangle = e^{\pm i a \int_{-\infty}^\infty\!dx\,
     [J(x,t)\hat\pi(x) + K(x,t)\hat\phi(x)]}|\Psi_0\rangle\;,
\ee
where the time-independent state $|\Psi_0\rangle$ is simply
$|\Psi_{\pm}\rangle$ with $a$ set equal to zero.  (This state is not
precisely the vacuum state of the field, but it resembles it closely
except near $x=0$, where it has been polarized by the unobserved
oscillator.)

$J$ and $K$ are the external sources which describe the spacetime
location of the information:
\ba\label{JK}
J(x,t) &=& {g\over2}\int_{-t}^t\!dx'\, F(x-x') r(t-|x'|) \nn\\
K(x,t) &=& {g\over2}\left( 2 r(t) F(x)  - F(x+t) - F(x-t)\right.\nn\\
& &\qquad\qquad\left. - (1+{2\over\pi}\epsilon)
       \int_{-t}^t\!dx'\, F(x-x') s(t-|x'|) \right)\;.
\ea
{From} \(JK) it is clear that in the late time limit, all the
information has propagated away from the origin.  In this late time
limit $ e^{-a^2 e^{-2\gamma t}} \to 1$, and the state of the field,
with the oscillator traced out, becomes pure.

Note that the behavior of $J$ and $K$ is not particularly sensitive to
the small amounts of initial field-oscillator entanglement that are at
issue in choosing \(init) instead of a direct product state.  The
operator exponent in \(source) is simply the projection, onto the field
sector of the total Hilbert space, of the time-evolved $\hat P$
operator:
\be
\hat P(t) = r(t) \hat P - M\Omega s(t) \hat Q
  +\int_{-\infty}^\infty\!dx\,[J(x,t)\hat\pi(x) + K(x,t)\hat\phi(x)]\;.
\ee
If the information initially in the oscillator is characterized by
$\hat P$, then the information propagation into the field will be
described (up to ambiguities near the origin) by \(JK). This clearly
shows that the information which was characterized by the initial value
of $P$ of the oscillator will become non-local in the late time limit.

\section{Discussion}

The evolution of the oscillator-field system shows that
the entanglement entropy is roughly constant until the oscillator
approaches its ground state.  Moreover we have shown that
the information is hidden in a very non-local way.

Both $J(x,t)$ and $K(x,t)$ are important in characterizing the state
of the field. $J(x,t)$ behaves very much like classical radiation from
a source, and implies that some of the information initially in the
oscillator propagates into the field in the same way that one might
naively expect.  In contrast, $K(x,t)$ possesses sharp spikes, whose
width is on the cut-off scale, which propagate away or decay in place.
The propagating spikes are evidently the couriers for the rapid
shedding of information that is associated with decoherence.  The
spike in $K(x,t)$ which remains at $x=0$ decays only on the
dissipative timescale.  This could be interpreted as describing
information which remains in the `quantum hair' of the oscillator, and
is only slowly radiated away.  It should be pointed out, however, that
our approximation in \(Simp) breaks down near the origin.

Furthermore, assigning the degrees of freedom near the origin to the
field or to the unobserved oscillator is to a large degree arbitrary,
since unitary transformations that alter such assignments can leave
the rest of the model essentially unchanged.  It could well be that
the information pool described by the non-propagating term in $K(x,t)$
is best considered as belonging to a `dressed' version of the
oscillator.  This type of ambiguity seems likely to be a common
feature of problems in which information loss and ultraviolet
regulators are closely related, since one does not expect a uniquely
specified ultraviolet cut-off to exist.  In fact, the best procedure
in such cases might be to search for an optimal kind of UV regulation
based on the location of information.

We can arrange to present the state of the field (with the oscillator
treated as unobserved) in the form of a Wigner function.  We can do
this by noting that (i) the field state is effectively a mixture of
pure states $|\Psi_\pm(t)\rangle$, and (ii) at any fixed time $t$, the
states $|\Psi_\pm(t)\rangle$ belong to a subspace of the field's
Hilbert space which may be mapped onto the Hilbert space of a single
harmonic oscillator.  The second assertion is justified by the inner
product Equation \(inprod), which is the correct inner product for two
coherent states with annihilation operator eigenvalues $\alpha_\pm =
\pm\sqrt{D(t)\over2}$.\footnote{By repeating the analysis of Section 2
starting with more general initial states, it is straightforward to
define a class of field states which represent, in the sense we are
considering, the full complex plane of coherent states, with complex
eigenvalues $\alpha$.}  Therefore, by applying to \(R4) and \(Simp)
the mapping
\be\label{map}
|\Psi_\pm(t)\rangle \to |\pm\sqrt{D(t)/2}\rangle_{osc}\;,
\ee
the state of the field may be described at any given time by a mixture
of single-oscillator coherent states $|\alpha_\pm\rangle_{osc}$.

Wigner functions for such mixed states are easily calculated.  The
function for the field at $t=0$ is single gaussian corresponding to the
ground state.  The Wigner functions which represent the state of the
field at several later instants are plotted in Figures 3.1 -- 3.3.
Figure 3.1 shows the state of the field very soon after $t=0$, as it is
just getting excited into a mixture of two coherent states. The two
overlapping gaussian peaks that can be discerned in 3.1 will separate
on the decoherence timescale.  (No rapid growth in energy is associated
with this sudden separation: it is only the inner products of the field
states, and not their energies, that fit the analogy with oscillator
coherent states.)  Figure 3.2 describes the field during the long
interemediate epoch of its evolution: a mixture of two well-separated
gaussians.  Finally, when the oscillator ends up in its ground state,
the field regains its purity.  The rapid oscillations near the origin
in Figure 3.3 are symptoms of quantum coherence.

These figures clearly show the loss and eventual restoration of quantum
coherence in the field.  In order to determine the degree of purity of
a harmonic oscillator, however, one must have some means of measuring
at least the gross features of its Wigner function.  In our case, the
observables which must thus be measured (in whatever combination) are
the non-local operators $\int_{-\infty}^\infty\!dx\,[J(x,t)\hat\pi(x) +
K(x,t)\hat\phi(x)]$ and its canonical conjugate.  Since these operators
are non-local on the dissipation scale, it would seem to be very
difficult to actually observe the asymptotic purity of the state of the
field.

This system can be thought as a specific example of Page's\cite{page}
alternative outcome of black hole evaporation. He suggests that
information might get out of black holes through radiation, and has
shown that the information might not show up in an analysis
perturbative in $M_{Planck}/M$.  Our system seems to behave in this
way, as in our case the information is not recovered until the very end
of the decay of the oscillator.  If we were to start the oscillator in
a {\it mixture} of the two gaussian states, instead of in the
superposition we have discussed, we would not notice any difference by
examining the field until the oscillator had relaxed to very near its
ground state.  How long this takes to occur, in our model, is dependent
on the initial state.

We should also point out that we do not find a strict relation between
the energy of the oscillator and the rate of information exchange (one
of the folkloric statements used as an argument for the loss of
coherence in black hole radiation).  Thus it might well be possible that
black hole evolution does preserve quantum coherence, but that it is
very hard, if not impossible, to recover all the initial information
in a set of local observations of the final state.

\section{Acknowledgements}

We would like to thank Andreas Albrecht for stimulating discussions.
J.R.A. would like to acknowledge the support of the Natural Sciences and
Engineering Research Council of Canada.

\eject
\openup 1\jot
\proclaim Figure captions.

\vskip 0.25 truein
\openup -1\jot
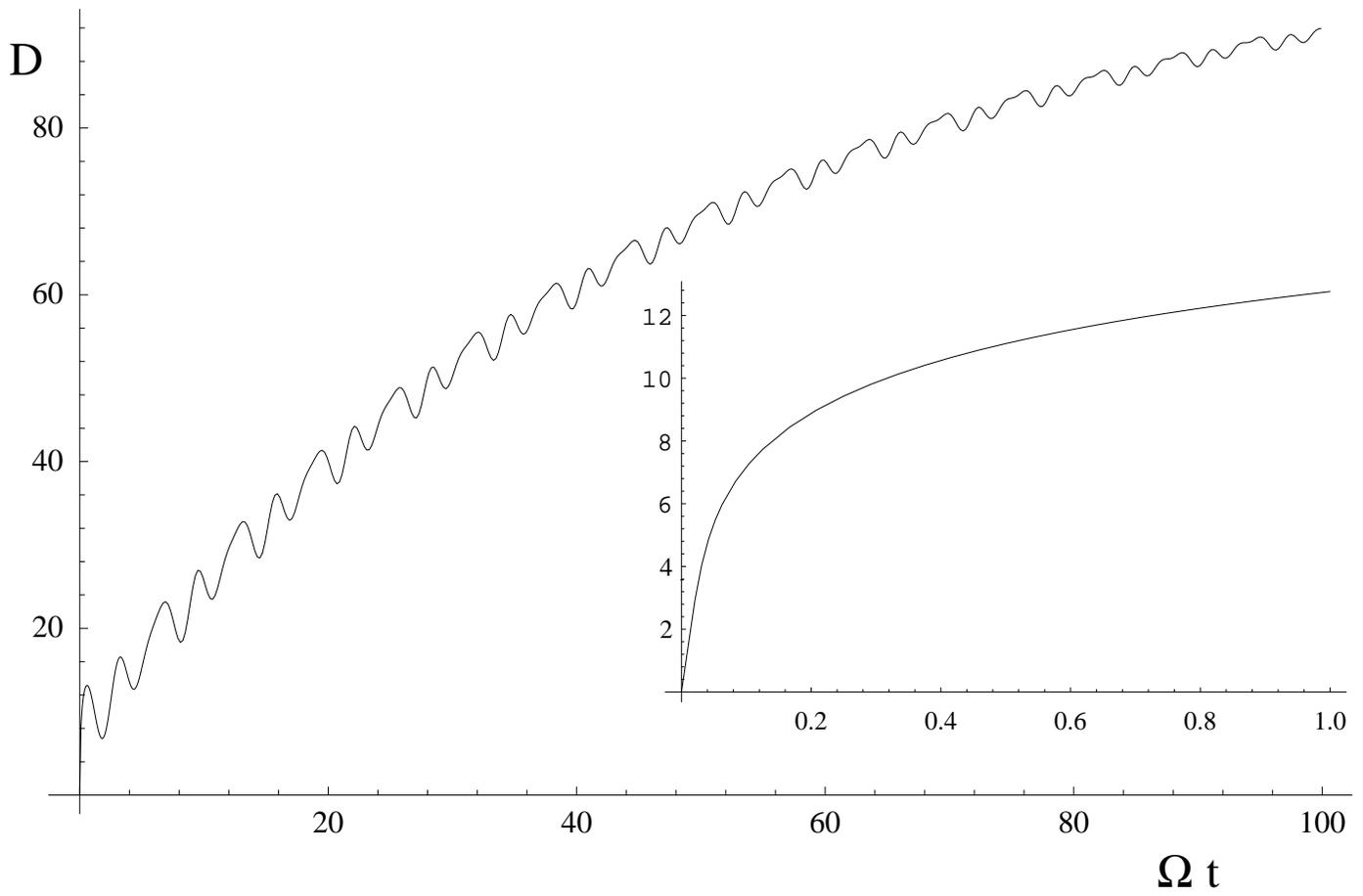
\noindent {\narrower\smallskip\noindent {\bf Figure 1.}
Behavior of $D(t)$ of equation (36), with the parameter $a$ set equal
to 10.  $D(t)$ is the term responsible for the suppression of the
interference term of the two-gaussian initial state; it increases
rapidly (on the decoherence timescale) and remains large.
\smallskip}     \openup 1\jot  \vskip 0.25 truein

\vskip 0.25 truein
\openup -1\jot
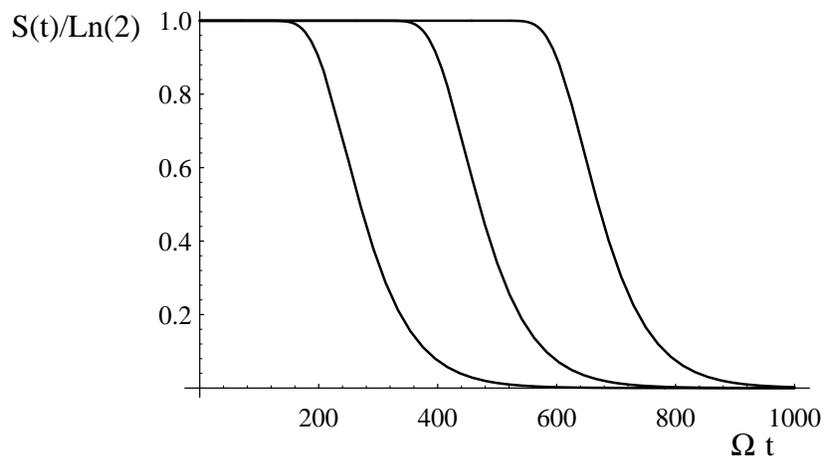
\noindent {\narrower\smallskip\noindent {\bf Figure 2.}
Evolution of the entropy of the oscillator for three initial states of
different values of the initial separation parameter $a$. The entropy
increases rapidly during the decoherence phase and remain esentially
constant until the oscillator enters the final stages of its decay
towards the ground state.  Since this occurs only when the oscillator
has lost all but a fixed amount of its initial energy, the length of
time before recoherence depends on the initial state.
\smallskip}     \openup 1\jot  \vskip 0.25 truein

\vskip 0.25 truein
\openup -1\jot
\noindent {\narrower\smallskip\noindent {\bf Figure 3.}
Evolution of the the degree of freedom of the field which is excited by
the oscillator ({\it i.e.}, the mode spanning linear combinations of
the time-dependent, non-locally excited states $|\Psi_\pm(t)\rangle$).
It starts in its ground state, is excited into a mixed state by the
oscillator, and approaches a pure excited state as the oscillator
decays to its own ground state.  The sharp oscillations in 3.3 are
witnesses to the purity of the field state at late times.
\smallskip}     \openup 1\jot  \vskip 0.25 truein

\end{document}